\documentclass[10pt,letterpaper]{article}

\usepackage{cogsci}

\cogscifinalcopy % Uncomment this line for the final submission 

\usepackage{pslatex}
\usepackage{apacite}
\usepackage{float} % Roger Levy added this and changed figure/table
\usepackage{graphicx}

\usepackage{cuted,tcolorbox}

\usepackage{amsmath}

\usepackage{bm} %for bold mathematical symbols

\newtheorem{principle}{Cognitive Process}
\newtheorem{definition}{Definition}

\newtheorem{theorem}{Theorem}
\newtheorem{lemma}{Lemma}

\title{MaTIC: a Mathematical Theory of Inferential Communication} %{How to Make a Proceedings Paper Submission}

 \author{Joan Llobera \\ Artanim Foundation \\  joan.llobera@artanim.ch
        \And
         Jordi Vallverdú \\ ICREA Acadèmia, Universitat Autònoma de Barcelona \\ jordi.vallverdu@uab.cat}

\begin{document}

\maketitle

\begin{strip}
\begin{tcolorbox}
\textbf{November 2021. } This paper summarises an attempt to  formalise inferential communication in a way that works for both humans and artificial agents. A typical application scenario would be social interaction between virtual characters and humans using Virtual Reality headsets and tracking hardware. 

Formally it attempts to build a non-stationary version of information theory, and develop which kinds of agents would be able to exchange information under this premise, trying to formalise in a mathematical model different notions well established in the cognitive science literature.

The reviewer feedback made us realise that to be convincing we needed to write a book if we wanted to address all the points raised. Since it is uncertain when we will have time to do this, we decided to publish this theory draft, even if not peer reviewed.

\end{tcolorbox}
\end{strip}

\begin{abstract}

The arrival of Immersive Virtual and Augmented Reality hardware to the consumer market suggests seamless  multi-modal communication between human participants and autonomous interactive characters is an achievable goal in the near future. This possibility is further reinforced by the rapid improvements in the automated analysis of speech, facial expressions and body language, as well as improvements in character animation and speech synthesis techniques. However, we do not have a formal theory that allows us to compare, on one side, interactive social scenarios among human users and  autonomous virtual characters and, on the other side, pragmatic inference mechanisms as they occur in non-mediated communication.

 Grices' and Sperbers'  model of inferential communication does explain the nature of everyday communication through cognitive mechanisms that support spontaneous inferences performed in pragmatic  communication. However, such a theory is not based on  a mathematical framework with a precision comparable to classical information theory.

To address this gap, in this article we introduce  a Mathematical Theory of Inferential Communication (MaTIC). MaTIC formalises some assumptions of inferential communication, it explores its theoretical consequences and outlines the practical steps needed to use it in different application scenarios.

\end{abstract}

\section{Introduction}

A key assumption of Shannon's mathematical theory of communication \cite{shannon48} was that communication was limited to random samples from a finite set of symbols, such as for example the alphabet, or zeros and ones. As Shannon, stated: \textit{Frequently the messages have meaning; that is they refer to or are correlated according to some system with certain physical or conceptual entities. These semantic aspects of communication are irrelevant to the engineering problem.} However, modern communication have obviated this limitation. It is common in modern communication systems to interact with chat-bots, virtual assistants, and different sorts of conversational agents.

More dramatically, modern virtual reality systems enable us to capture, stream and render full three dimensional representations of human actors, across the auditory and the visual domains. Alternatively, such performances can also be created with digital production tools, on a labour-intensive practice by character-animation specialists. We can therefore create, store and replay animations of cartoon or realistic characters. Moreover, we can render interactive performances  using Virtual or Augmented Reality headsets, thus recreating a communication scenario that, despite being technologically mediated, can be quite similar to spontaneous natural communication.

For such a communication scenario to match our expectations,  artificial cognitive agents must infer (or appear to infer) the meaning of a certain message and, from a combination (or an abstract representation) of previously recorded performances, generate a response that is appropriate to the message received, and that is (or appears to be) aligned with the communication intentions of the cognitive agent.  Several techniques can be adapted for this purpose, either from natural language processing, or from artificial intelligence literature focused on video game characters.  These techniques, though, are often in stark contrast with the way such mechanisms occur in humans. We lack a communication theory comparable to the precision of Shannon's work to characterise precisely inferential communication, as well as to quantify the benefits and drawbacks of different approaches to create artificial agents capable of generating interactive behaviour consistent with these principles.

To address this gap, in this article we introduce a Mathematical Theory of Inferential Communication (MaTIC).  The article is structured as follows.
After this introduction, we review how inferential communication is described in cognitive pragmatics and review broader cognitive principles.  We review notions such as predication, natural categories, conceptual metaphors, and competence acquisition.

Next, we briefly review how cognitive and computational neurosciences characterise the embodiment of such principles. We review the concept of massive modularity as a cognitive organisation principle and discuss how such functional modules inter-relate in inhibitory and excitatory networks. 

We then summarise the previous picture in  a formal construct called the general cognitive module (GCM).  We argue it summarises how cognitive sciences characterise the building blocks of embodied cognition. We also try to  
 illustrate the generality of a GCM we show how well-known computational models for autonomous agents can be interpreted as particular cases of such a construct.

We then turn to our main case of interest:  how cognitive agents formed of networks of GCMs can learn to interpret the actions of other  agents and generate appropriate responses.  For this purpose, we introduce three cognitive principles for inferential communication that we label $I$, $S$ and $T$. We  interpret these principles as mechanisms that shape  the topology of connections among different GCM. 
These guide  \emph{competence acquisition} in an agent, and our proposal is that such principles are as valid for humans as for artificial agents.

A particularity of this theory is that it tries to reconcile the assumption that subjective inferences, as they occur in inferential communication, emerge from embodied mechanisms, with a formalist approach. We try to illustrate how these two aspects make a quite compelling picture in a domain where subjective inferences and a formalist approach converge: mathematical foundations as depicted from the perspective of embodied cognition. In this context, we show that the resulting theory under specific assumptions produces a set theory that, contrary to  the axiomatic assumptions of ZFC set theory, the set theory assumed by default, is consistent with the picture of embodied cognition as a foundation for mathematics.

Finally, we discuss whether such a model allows adapting the rich tool set of  information theory to characterise information exchanges between two inferential agents, how it compares with different proposals that try to define cognitive systems with general computational principles, and  future research directions, both theoretical and practical.

\section{Inferential communication}

Contrary to the code model \cite[]{shannon48}, in which a communicator encodes a message to be decoded by an audience, the \emph{inferential} model of communication considers that \emph{communicative actions}, both verbal and non-verbal, should be understood as the expression and recognition of intentions: the communicator provides evidence of his intention to convey a certain meaning, and the audience infers the meaning on the basis of the evidence provided. This evidence can be something said together with some gestures and voice inflexions, but it can also be non-verbal, such as pointing, smiling, approaching, or any other action that the audience is susceptible of associating a meaning to it. The association of the evidence with a meaning is strongly influenced by the social and material context.
If classical information theory was largely inspired by the image of two agents coding and decoding signs through a telegraph wire, the inferential model of communication can be exemplified by the communication of two fully embodied agents sharing a three dimensional environment portraying in real time bidirectional behaviour and speech with the purpose of exchanging information and intentions related with this information. This model has never been formalised or tested in quantitative and rigorous terms such as information theory has. On the other side, it can give simple explanations to quite complicated phenomena. For example, in this view the reason why we are constantly playing with different layers of meaning when we communicate is simply for communicative economy. To give a better picture of this theory, we summarise some of the notions on which it is based:

\begin{enumerate}
    \item \textbf{Language games}\cite{wittgenstein1953pit} characterised  everyday communication as a \emph{language game}.  To do so, he outlined a large number of examples on how the meaning of words and actions are determined by how we use them, in what context and with what intention. Therefore, we can define   the class  of \emph{Language Games} as all situations in which several people (or things like virtual actors) can perform \emph{communicative actions}.  
    \item \textbf{Natural Categories}
The notion of Natural Categories \cite[]{rosch1996frs} has been widely adopted in cognitive sciences, ranging from cognitive psychology to applied linguistics. Essentially, natural categories are the basic element that allows us to categorise the world and organise mentally cognitive knowledge. Formally speaking, they were generally identified with fuzzy sets \cite[]{zadeh1965fuzzy}. 
  This implies there are better and worse examples of each category, which it is often named as the level of prototypicality. Obvious examples of this can be found in colours: in face of a spectrum of different red samples we will attribute a ``typical red" example or a ``not so good red" example to different samples.

\item \textbf{Conceptual Metaphors}
In cognitive linguistics, the notion of conceptual metaphor as used by \cite{lakoff1980mwl} builds upon the idea of natural categories. In this view, conceptual metaphors are mappings between domains that give a richer meaning to how we communicate and think. Usual metaphors such as \emph{Life is a Journey}, or categories of different cultural systems, such as  the category of \emph{Women, Fire, and Dangerous things} \cite{lakoff2008women}, to cite only the most famous examples, rely on the notion of Natural Category. 
 This interpretative framework has had numerous developments. For example, in arguing that basic mathematical intuitions rely on conceptual metaphors that are derived from experience, and that it is this which gives a lot of what we call mathematical understanding, independently of the demonstration process \cite[]{lakoff2000mc}.

\item \textbf{The Context of an action}

  \emph{Communicative actions} related with interpersonal communication have  a context which can include factors such as: 
  part of the physical situation, but also the cultural background assumed to be shared, the gender, social relation of the speakers and in general anything that contributes implicitly to determine what is meant by a certain communicative action.\footnote{The role of  contexts in natural communication can be traced back to 1923 in the work of Malinowsky, but see chapter 2, section 1 in \cite{kramsch1993cac}, or  the entry \emph{context et situation}  in \cite{houde1998vsc}}  A typical example of this is irony. If someone says \emph{I am very happy}, his assertion can be interpreted as meaning to express his happiness, or exactly the inverse, his extreme sadness. This will depend not only on how this is said, but also according to what we know about the person from what he did in the past. Some elements of the present situation can also have an influence in what is interpreted from what is said. Therefore, in natural communication agents must control very carefully the context in which \emph{communicative actions} are performed. Otherwise it will be difficult to ensure the meaning intended is appropriately conveyed.

\item \textbf{Intentions}
A different aspect of interpreting events when performed by actors is that we associate \emph{intentions} to them.
If we see someone in a certain situation or conversation, especially if we know him well or he is in a familiar socio-cultural context, we can spontaneously say what he will probably do or say, or at least a small set of different options, according to the situation, taking into account previous experienced contexts. This is a general cognitive ability of humans called \emph{intention attribution} or \emph{mind-reading}, which some impaired people seam to lack \cite[]{baron2000understanding}.

For example, consider the expression: \emph{This paper is white.}
 Depending on the context in which it is said, the same text can convey a meaning similar to \emph{I’m surprised because in his exam he did not write a word}, or \emph{Fantastic, finally I found a blank paper and I can draw what I was wanting to} or \emph{so, you finally did not fill in the papers of our divorce, did you?}. The process by which we go from \emph{what is said} to \emph{what is meant} is determined not only by the meaning of the words and in what situation the communicative action is performed, but also by  implicit assumptions in everyday interaction. All these are part of the context.

With the previous notions, we can summarise inferential communication  in non-mathematical terms with the  \emph{Principle of Cooperation} \cite[]{grice200lac} and the two principles of \emph{Relevance Theory} \cite[]{sperber2004rt}.
 
\item \textbf{The Cooperative Principle} 
%\subsubsection{The Cooperative principle}
\label{coop-principle}
Grice's \emph{Cooperative Principle} was the first principle of communication proposed, and it would involve any everyday linguistic interaction --even non-verbal. The clearest explanation might be introduced by it's proponent \cite[]{grice200lac}):
\begin{quote}
    Our talk exchanges do not normally consist of a succession of disconnected remarks, and would not be
rational if they did. They are characteristically, to some degree at least, cooperative efforts;
and each participant recognises in them, to some extent, a common purpose or set of
purposes, or at least a mutually accepted direction. This purpose or direction may be fixed
from the start (e.g., by an initial proposal of a question for discussion), or it may evolve
during the exchange [...]. We might then formulate a rough general principle which participants will be expected (ceteris paribus) to observe, namely: Make your conversational contribution such as is required, at the stage at which it occurs, by the accepted purpose or direction of the talk exchange in which you are
engaged. One might label this the \emph{Cooperative Principle}.
\end{quote}
A classical example adapted from \cite{grice200lac} will frame better the problem. Consider this scenario:

\begin{quote}
A is standing by an obviously immobilised car. \\
 B approaches him. \\
 A says: I am out of petrol. \\
B says: There is a garage around the corner. \\
\end{quote}
In the previous example, the last reply of B would be a consequence of the previous exchanges. 
To formalise such a mechanism we will need a definition of implication ($\Rightarrow$) general enough to include these \emph{implicatures}.

\item \textbf{Relevance Theory}
Sperber and Wilson's Relevance Theory \cite[]{sperber2004rt} develops two basic principles to explain the inferential model of communication. The first principle of relevance theory or cognitive principle, states that \emph{human cognition tends to be geared towards the maximisation of the cognitive effect}, and this cognitive effect is quantified with a  relevance measure. This implies that, given a certain context, a communicative action will be performed in order to maximise the relevance that the receiver will associate to it. 
The second principle of relevance theory or communication principle adds an extra layer. It states that \emph{every ostensive stimulus conveys a presumption of its own optimal relevance}. This is best shown with an example. If two people, Alice and Bob, are chatting in Alice's living room, Bob can put his empty glass in Alice's line of sight to suggest that he might want more water, but this is only exploiting Alice's cognitive tendency to maximise relevance. Bob still has to show you that he put the glass there \emph{on purpose}, that it was his intention, and therefore that it is worth processing this stimuli in Alice's cognitive system, which is something not trivial in a world of stimuli competing for scarce cognitive attention.
The previous implies that a communicative event is not one in which a maximum amount of information is given, but rather one in which a maximally \emph{relevant} information is given. 
It is somewhat a principle operating at a second level: it implies any event triggered will aim at giving the minimal information but with the maximal relevance. At reception, it also implies that determining the goal(s) conveyed by a certain action should assume optimal relevance. But to be able to quantify such relevance, agents need not only to determine causal relations, but also have the skills to determine the underlying goals of a certain behaviour.

\item \textbf{Competence Acquisition}
 The ability to associate intentions with human behaviour is developed by sustained social interaction from the very first stages of baby learning \cite[]{halliday1985saw,tomasello2003clu}. Competence acquisition in language and communication seems to be based on imitation combined with the development of mental schemes. Language competence would come progressively by associating the performance of different chunks of language with real-world contexts and consequences. Progressively the learner would use these chunks of language and their combinations.  Therefore, the connection between sounds, grammar,and meaning is not an universal process, but a  social and situated one \cite{kendrick2020sequence}. According to Kendriks, there appear to be strong universals of interactive language usage,
namely the sequencing of social actions across neighbouring turns, and this allows to identify the universality of the sequence organisation observable in informal human conversational interaction.  There is some experimental evidence this describes accurately how kids acquire grammatical knowledge \cite[]{bannard2009modeling}.

 This also seems to correspond with the old notion of \emph{internalisation} as a means to acquire communicative competence \cite[]{vygotsky1964tal}. For example, one learns how to ride a bicycle by seeing it and trying it, independently of an understanding of how it works: you internalise what to do to get a certain effect.  Similarly, one learns to use a word by determining in what context and with what purpose someone else used it, and then shifts to use it in order to obtain a certain effect.
The term \emph{internalisation} therefore refers to  the process by which, given a set of examples on a particular domain, we construct mental categories that abstract particular examples and give us the skills to operate, in  a variety of circumstances, on that particular domain. In this picture, kids and adults acquire novel mental categories by internalising from few examples given  \cite[]{Vygotsky1986}.  
 \end{enumerate}

The previous notions give a (quite brief) account of the general assumptions of inferential communication.  To summarise this section, we are assuming that in \emph{Language games} the meaning of a \emph{Communicative action} is determined from how it is used. We call the meaning an \emph{interpretation}. The \emph{interpretation} of an action requires associating intentions or goals to it. To determine these an agent needs to categorise the action with a large set of  \emph{Natural Categories} interrelated with \emph{Conceptual Metaphors}. The agent also needs to monitor the \emph{Context} in which each \emph{Communicative action} will occur. 
\emph{Causal inferences} are core mechanisms of the agents to do so. These inferences take the form of \emph{Conversational implicatures} and they are based on \emph{Relevance principles} shared among the agents.
All these skills are acquired through interactive mechanisms of imitation and cognitive development in a process called \emph{Internalisation}. %, enriching mental representations through interactive development. 

\section{A general cognitive module}
\label{embodied-model}

\label{cog-module}

\label{embodied-reasoning}

\subsection{The model}

 Sperber convincingly argues that the specifics of how we draw cognitive inferences is best explained through massive modularity , i.e., the existence of a massive amount of interrelated functional modules \cite[]{sperber1996ecn}.  An example he draws is the use of masks in primitive rituals, and how the functional module specialised in faces is here used in an altered way. In this sense, the evolution of culture is to be understood as how  brain plasticity  shapes such functional modules to better fit their use among social agents. Building on Sperber's vision, we assume that in inferential communication an agent is composed of a massive amount of interrelated functional modules. 
 
To formalise inferential communication among humans and artificial agents, we need to characterise such functional modules, and do so in a way that works both for humans and for artificial agents. To do so we introduce a formal model that we call a \emph{general cognitive module} (GCM). Despite it is non-linear and difficult to treat, it has the advantage of being quite general. 
Our working hypothesis is that such a structure is general enough to be used to implement the different cognitive mechanisms previously outlined, in the same way that combinations of NAND logical gates can be used to compile any digital algorithm. 
Despite  we do not provide a proof of such an assumption, we do try to show that under certain assumptions such a model fits with well understood cognitive systems.

\begin{figure}

\begin{center}
\includegraphics[width=8cm]{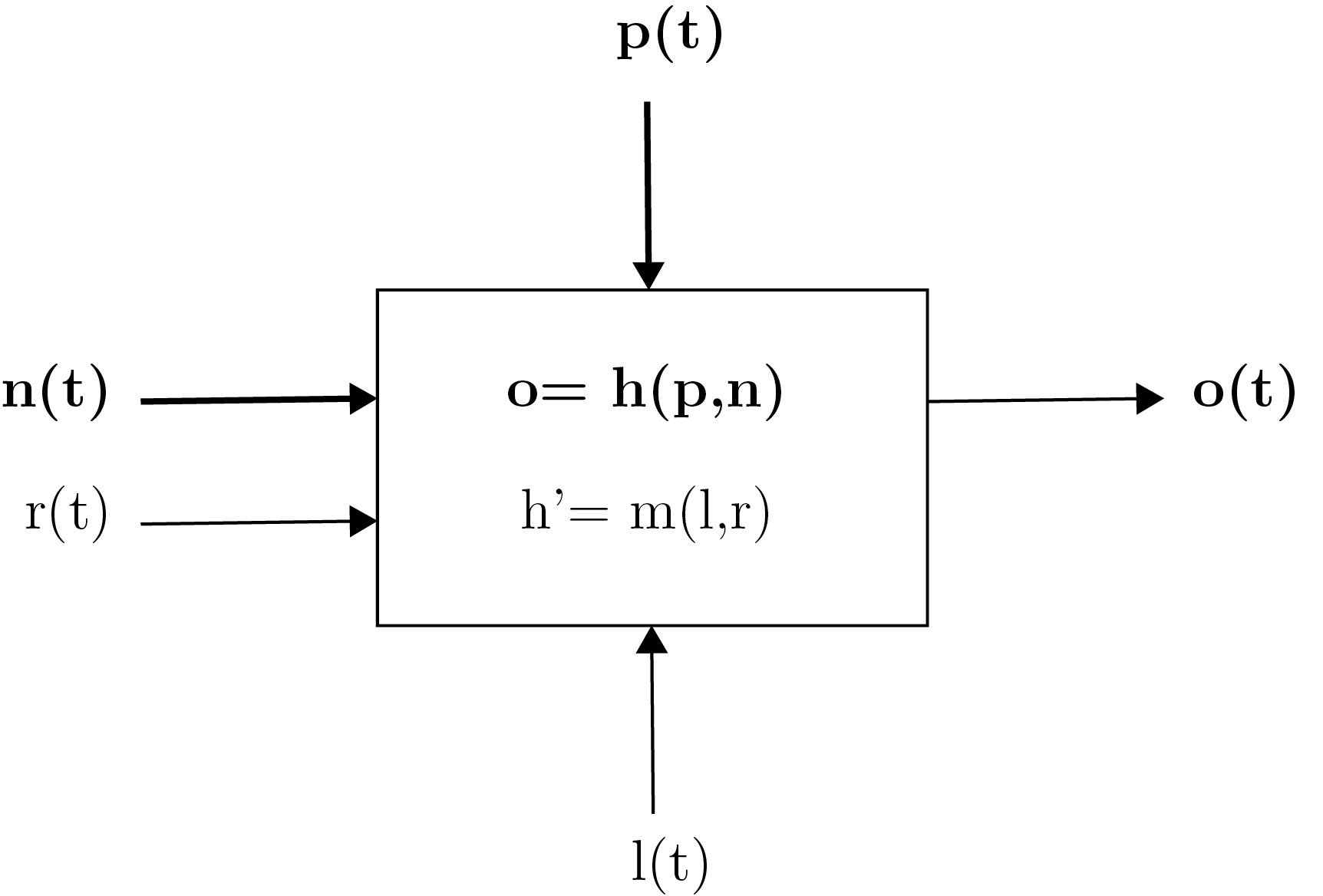}% This is a *.eps file

\end{center}
\caption{ The different elements of a general cognitive module. On the higher half, in bold, the fast pathway. On the lower half, in plain text, the slow pathway. }\label{fig:1}
\end{figure}

A GCM is composed of two pathways, the fast one and the slow one.  The fast pathway is defined as a transfer function that  is an invariant, non-linear filter with a response function $\boldsymbol{h}$. This pathway is defined by the equation $\boldsymbol{o(t) = h(p(t),n(t))}$, and it is composed of:
\begin{enumerate}
    \item $\boldsymbol{o(t)}$: A stochastic output arriving through from electrical synapses. In the case of a neuron this is the axonal output firing action potentials. In the case of a piece of brain tissue this can be a statistical distribution of action potentials.
    \item $\boldsymbol{p(t)}$: An array of stochastic inhibitory inputs . For example, in the case of a neuron or a piece of brain tissue these will be inputs from inter-neurons arriving through from electrical synapses. This signal would define whether the output is \emph{possible} or not.
    \item $\boldsymbol{n(t)}$: An array of stochastic excitatory inputs . For example, in the case of a neuron or a piece of brain tissue these will be inputs from pyramidal neurons arriving through from electrical synapses. This signal would define whether a possible output would \emph{necessarily} occur. 
\end{enumerate}
The slow pathway updates the previous response function. 
It uses the following signals:
\begin{enumerate}
    \item $r(t)$: An array of reward inputs.  For example, in the case of a neuron or a piece of brain tissue these will be inputs from chemical synapses.
    \item $l(t)$: A learning input which, for simplicity, we assume can activate or disable learning. In a neuron or brain tissue this would be chemical inputs from the neuroglia, enabling or disabling metabolic plasticity. 
    \item  $m(l,r)$ A metabolic response, which determines the update of the transfer function: $h'=m(l(t),r(t))$. When considering  $h'$ we may also want to consider the previous transfer function $h$ as a constant in the metabolic response $m(l,r)$.  
\end{enumerate}
Overall we can think of general cognitive modules (GCMs) as a scale independent, simplified representation of a computational mechanism for cognitive processing. As such, a correspondence can be established with a biologically realistic neuron, but also with  a fragment of brain tissue, to a unicellular animal, or even to more sophisticated natural cognitive systems. It can also be matched to  an artificial neuron, or to different machine learning architectures.

\subsection{GCMs in natural cognition}

To illustrate how a GCM fits with the idea of  a human brain analysed as a massive amount of GCM modules, we show how it fits with cognitive and computational neuroscience. Neuroscience textbooks give us some hint on how a brain tissue might instantiate a functional module in practice. The nervous tissue constantly balances inhibitory and excitatory signals \cite[]{buzsaki}. Typical examples of excitatory pathways are pyramidal cells in the brain cortex, which transmit signals at long distances. Typical inhibitory signals in the brain cortex are inter-neurons, which are shorter, inhibitory pathways, and more abundant than excitatory ones. In a GCM module, inhibitory  pathways correspond to  $\boldsymbol{p(t)}$, and excitatory pathways correspond to  $\boldsymbol{n(t)}$.

We do not have a clear understanding on how information is encoded in the central nervous system. Despite a common assumption in modelling approaches is to study neural firing rates, it has been shown that temporal patterns also play a major role in coding perceptual information (see, for example, \cite{Gutig2006}). We therefore need to consider different encoding options are possible.  However, what we do know that in brain tissue outputs are determined by, first, not having inhibitory inputs and, second, having enough excitatory input to stimulate an output. In a GCM, outputs are defined by  $\boldsymbol{o(t)}$, and the relation between input and output determines a,  comparatively, fast, non-linear transfer function, denoted as  $\boldsymbol{h(p,n)}$.

Opposite to the fast pathways, chemical synapses are complex chemical reactions, with a slower processing time, and determine the metabolic consequences of excitatory and inhibitory input.  

The dopamine system, along with several other chemical signals, can inhibit or activate the functional plasticity of the tissue (in a GCM, this corresponds to $l(t)$). Other signals will make the response of the cell change to respond better or worse to fast stimuli (in a GCM, this corresponds to $r(t)$).Chemical synapses can also trigger proteine transduction from gene expression, which ultimately change the shape of cell and  the non-linear transfer function between inputs and outputs. In a GCM, this corresponds to $m(l,r)$.
The previous picture is quite general, and it is consistent with brain areas with very different functions, different connectivity, and different stimuli-response electrical patterns  (for example, between the cerebellum, the prefrontal cortex and the basal ganglia).

Another role of chemical communication, beyond synapses, is to shape network connectivity, mostly through chemical gradients spread across brain tissue.
Arguably, the mechanisms that are behind the establishment of network connectivity are what determine the brain's extreme parallelism, high level of recursivity patterns, as well as specific recursive circuits such as, for example, thalamo cortical loops for multimodal integration, or  cortico-basal and ganglia-thalamo-cortical loops related with the dopamine reward system. 
 Moreover, the capacity of the brain to change function and connectivity is not only related with neural activity. It also involves  the support system. For example, astrocyte signaling regulates plasticity \cite{ota2013role,haydon2015astrocytes}. Other  phenomena, such as brain cell migration in early neurodevelopment \cite[]{tierney2009brain}, also has an important role in brain plasticity. The mechanisms supporting network connectivity are not defined in the GCM. We will later introduce some principles to guide connectivity between GCM modules. However, before doing this, we turn towards artificial systems, to show how GCM modules can formalise, also, computational agents devoid of any biological realism.

\subsection{GCMs in artificial systems}
\label{specific-cases}

To illustrate the generality of this formalism we now discuss some specific, well known, cases, from the perspective of such a model.

\subsubsection{A Classic Communication System}

If we target at a classic communication system \cite[]{shannon48}, we can design a receptor as a single instance of the General Cognitive Module previously outlined. This can be done quite simply by doing the following additional assumptions:

\begin{enumerate}
    \item The slow pathway is completely static ($h'=h$), Reward inputs are null ($r(t)=0$) and Learning inputs are also null ($l(t)=0$).
    \item  The excitatory input array ($\boldsymbol{n(t)}$) becomes, simply the baseline signal, i.e., the signal transmitted after the radio-frequency filter and the demodulation step. 
    \item The inhibitory inputs ($\boldsymbol{p(t)}$) are a digital switch, encoding the connectivity in the graph of possible symbol transitions
    \item The transfer function ($\boldsymbol{h(t)}$) is a simple array of linear and invariant matched filters responding to the excitatory input, turned on or off by the inhibitory signal 
    \item The fast pathway has a stochastic output  ($\boldsymbol{o(t)}$) defined by the modulation technique (defined in the matched filters) the coding (defined by the possibility signal in the inhibitory inputs), and the received input (defined in the necessity, excitatory input).
\end{enumerate}

\subsubsection{A Video game Character}

A typical way to create an interactive character in  video game industry is to combine a behaviour tree, which makes the decisions, together with a hierarchical state machine which takes care of blending the relevant animations. 
A behaviour tree is defined as a tree-like graph, where different control nodes will select which branches are active. Some of these control nodes will evaluate binary conditions, and some will simply select leaf nodes sequentially, or randomly.
Leaf nodes, in turn, will execute a behaviour, and return a binary outcome.
Executing a behaviour means a virtual character being rotated or displaced in a 3D environment, often also triggering a kinematic animation, often through a hierarchical state machine. The transition graph of the state machine will define the possible transitions between animations. Additional parameters will define when these transitions can be triggered.
Skeletal animation techniques \cite[]{magnenat1988joint} will be used to render the body mesh. \footnote{ Different techniques for facial animation are possible, but for the purpose of this analysis we will assume there are no facial expressions involved.} 
Such an architecture can also be considered a particular case of our GCM. Below we outline a possible implementation:

\begin{enumerate}
    \item The input from the environment ($\boldsymbol{n(t)}$) will simply be a set of binary conditions, perceived from the virtual environment inhabited by the character. 
    \item The inhibitory input ($\boldsymbol{p(t)}$) will be defined by the animator. It will define the graph of possible transitions, i.e., when an animation can transition to another animation.
    \item The output     ($\boldsymbol{o(t)}$) will have three parts:
    A) a skeleton, i.e., a graph specifying the connectivity between bones, and the size of each.
    B) the 3D position of the root node
    C) a set of 3D rotations,  corresponding to each node in the bone topology,  generally expressed as quaternions.
    \item the transfer function ($\boldsymbol{h(t)}$) can be thought as a set of impulse responses of linear invariant filters, one for each kinematic animation stored. The selection of each impulse response will depend both on the possibilities enabled by the transition graph.
\end{enumerate}

Of course, it would also be possible to consider the decision and animation parts as separate GCM modules, where the output of the first is connected to the input of the second. However, for this purpose we need some kind of logic that helps us connect the different modules, which we will address in next section. Before this, we illustrate with an example how the slow path of a GCM can be used.

\begin{figure}
\includegraphics[width=8cm]{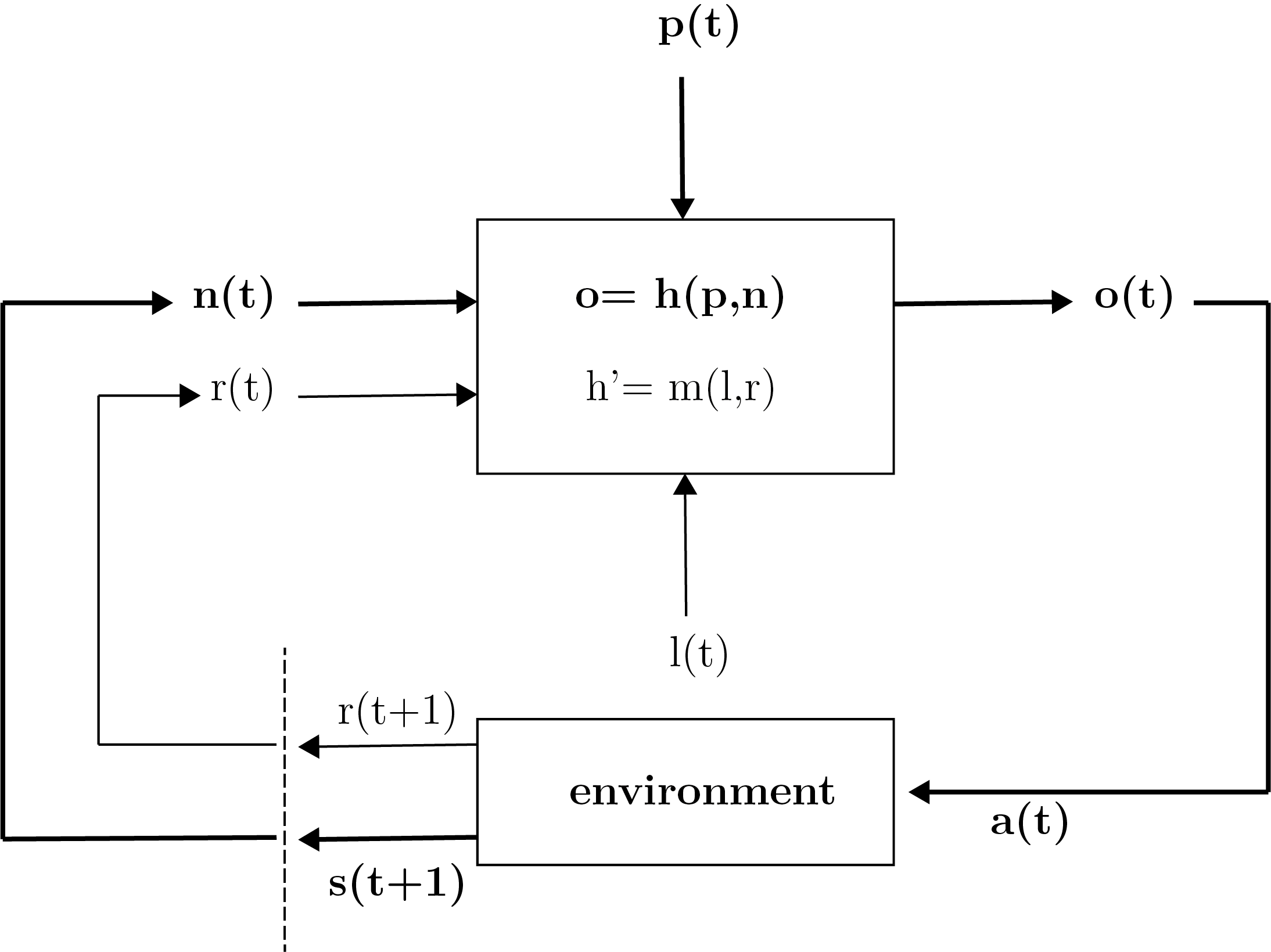}
\caption{Our diagram for a  GCM integrated in the general reinforcement learning framework }

%developed in \cite{sutton2018reinforcement} }

\label{fig:2}
\end{figure}

\subsubsection{Character animation based on reinforcement learning}
Some of the most impressive demonstrations for computational cognitive systems are
based on Deep Reinforcement Learning. Recent systems have demonstrated super-human features in classical games such as go or chess. However, to further illustrate specific cases of the GCM introduced we choose to analyse an example closer to our main interest: the use of reinforcement learning for physics-based interactive character animation \cite[]{peng2018deepmimic}\footnote{A general introduction and demonstration videos can also be found in https://xbpeng.github.io/projects/DeepMimic/index.html}. Such a system uses the same structure than a general reinforcement learning problem \cite[]{sutton2018reinforcement}. In figure \ref{fig:2} we show how our GCM fits in a RL framework. Compared to the previous section, the main changes are:

\begin{enumerate}
    \item The slow pathway is not static ($h' \neq h$). Reward inputs ($r(t)=0$) are defined from the connectivity of the agent with the environment. Learning inputs ($l(t)=0$) will be an inverted step function: one for the training phase, null for the animation synthesis step.
    \item In the fast pathway, the transitions between animations are not determined by an external state machine. Therefore the inhibitory input ($\boldsymbol{p(t)}$) is not defined. The transfer function ($\boldsymbol{h(t)}$) is not anymore a set of linear invariant filters, one per each animation. Rather, it is a non-linear function with internal recursivity: triggering an animation means increasing the importance of an attractor in a dynamic system. The stable state of the attractor is the synthesis of an animation. The excitatory inputs ($\boldsymbol{n(t)}$)  are not only the binary  variables that previously triggered a given animation, but also dynamic inputs such as the last pose of the character, as well as arbitrary parameters, typically target positions and orientations of skeletal nodes such as the root, a feet or a hand.
    
\end{enumerate}

The previous examples show how we can consider a GCM as a generalisation of diverse specific, well known, cognitive systems.  However, as already stated, we do not know anything about the connectivity and dynamic exchanges between different GCMs and  agents (made of GCMs). We address this in next section.

\section{A Formalism for Inferential communication}

In this section  we introduce formal definitions to relate the events and the pragmatic inferences characteristic of  inferential communication. Theorem proofs are in the appendixes.

\subsection{Events and Predicates}
First, we formalise events and predicates associated to them. 

\begin{definition}  A specific \textbf{event} is denoted with letters such as 
`$a, b, c, ...$'. A generic event is denoted with the letters `$x, y, z$'.
We may use sub-indices if further variables are needed, such as $x_1$. We assume an event $e$ has a beginning in time $t_e$, and a duration $d_e$. 
\end{definition}

Notice that these events occur in a specific time period, and as such they are not the abstract, instantaneous events, found typically in probability theory.

\begin{definition}  A \textbf{property} is denoted with letters such as `$A, B, C, ..., P, Q, ...$'.  We may also sub-indices if further variables are needed, such as $P_1$. 
\end{definition}

We also introduce an epistemic definition of logical quantifiers relative to the notion of \emph{a cognitive system}. 

\begin{definition}
A \textbf{cognitive system} is defined as the combination of one or more GCM instances connected between them through at least some of their inputs and outputs. 
\end{definition}

\begin{definition}
 The symbol `$\exists^{st}$'
 denotes the existence of an event external to a cognitive system.  We read it as \textbf{standard existence}. For an event `$e$' we write `$\exists^{st}e$'. For a property `$P(e)$' we write `$\exists^{st}P(e)$'.  An event `$i$' whose existence is represented internally by the cognitive system, but not necessarily associated with an external event, it is denoted as `$\exists i$'. Conversely, a property associated to such an event is denoted `$\exists P(i)$'.
\end{definition}

To denote an event with standard existence we call it  a \textbf{standard event}. It is important to notice that this does \emph{not} correspond to the classical idea of events as atomic elements, and properties as equivalent to predicates defined as sets of atomic elements. Here \emph{standard} events are associated to the input of a cognitive system, and properties are associated to its outputs. As such, if the outputs of a cognitive system  is connected to the input of a second cognitive system, the property will become an event \emph{from the perspective of the second cognitive system}. The transition from standard existence to general existence will be discussed in detail in later sections.

Consistently with the previous definition, we also introduce the standard universal quantifier:

 \begin{definition}
 The symbol `$\forall^{st}$'
 denotes \textbf{all standard events}.  The predicate `$\forall$' denotes all events. 
\end{definition}

The epistemic nature of the standard existential and universal quantifiers is particularly visible when the cognitive system under consideration is an entire agent. In this case, `$\exists^{st}$' denotes existence of an event \emph{in the world}, as perceived by an agent. The event is the external stimulus, and the property is the perception. Different from this, `$\exists$' denotes existence in more abstract terms, as is done in usual mathematical logic. For example, it could refer to a hypothetical event, independently of whether it is associated to an external stimuli or not.

A class of properties that will become very important are those that can also be considered sets:

\begin{theorem}
The outputs of all non circular cognitive systems are Sets. 

\label{predication-theorem}
\end{theorem}

\begin{definition}
\textbf{Predicates} are properties that are also sets (i.e., that are the outcome of non-circular cognitive systems). 
\end{definition}

Since predicates are also sets, they are close to the sense of predicate given in predicate logic and first order logic. We can therefore introduce the usual operations of set theory for  predicates:

\begin{definition}
The operator `$\in$' denotes a relation between an event and a predicate. It reads as \textbf{belongs to}, or \textbf{in}.  For example,  $ a \in  P $ denotes `$a$  satisfies $P$' (i.e., $a$ is associated with predicate $P$). We also denote this relation as $P(e)$.
We may also consider fuzzy belonging, denoted as $\in_{\alpha}$, with $\alpha \in [0..1]$. We also use $\notin$ to denote the negation of $\in$. 
\end{definition}

\begin{definition} To combine predicates we use the the union ($\cup $) and intersection ($\cap$) operators. To denote two predicates are equivalent we use `$\equiv$'. The opposite is denoted as `$\not\equiv$'.

\end{definition}

We now have a complete formalism to combine and compare \emph{predicates}.
\subsection{Causal Implicatures}

To formalise \emph{causal implicatures} we introduce event sequences:

\begin{definition}
 A \textbf{chain of events} is denoted with the `$.$' operator. For example `$a.b.c$' denotes a chain of three events. The only constraint introduced by this operator is that the occurrence of the events satisfies a partial order, i.e.,  that $a$ starts occurring before or at the same time than $b$, etc. Therefore, if $t_a$ denotes when $a$ starts occurring,  `$a.b.c$' implies $t_a \leq t_b \leq t_c $.
\end{definition}

 There are no constraints regarding when events end (i.e., since events have a duration, $a$ may finish occurring after $b$). From the previous, we can define:

\begin{definition}
The \textbf{context} $x$ of an event $y$ is denoted as  $y^*$. It is defined as a chain: $y^* \equiv x_1 . x_2 . ... . x_n$. It does satisfy the condition that $t_{y^*} \leq t_y$.
To denote a context $x$ as perceived by an agent we write $X$, i.e., the predicates associated with the events in $x$.
\end{definition}

\begin{definition}
	The \textbf{implied cause} $x$ of an event $y$ is denoted as  $y^{**}$.
	It does satisfy the condition that $t_{y^{**}} \leq t_y$.
	
\end{definition}

\begin{definition}
The relation between an event $e$ and its implied cause $e^{**}$ is called a \textbf{causal implicature}, and we denote it as  `$e^{**} \Rightarrow e$'.

\end{definition}

\begin{theorem}
For a given event $e$, it is always possible to build a cognitive system that can determine the most likely implied cause using an information-based metric such as surprise minimization.
\end{theorem}

\begin{definition}
An \textbf{inferential agent} is an agent that builds causal implicatures based on information-based metrics such as surprise minimisation.
\end{definition}

 In this picture, an inferential agent trying  to interpret a communicative exchange will build pseudo-complemented semi-lattices to establish causal implicatures and select the construct that minimises  uncertainty.

\begin{theorem}
An information exchange between two inferential agents with goals is not stationary.
\end{theorem}

The previous result shows in what sense this theory generalises  classical information theory: contrary to traditional information theory where events are part of a  stochastic stationary process, here the probability of occurrence of an event will vary across time.

\subsection{Cognitive processes for competence acquisition}

\label{internalization-principles}

An important aspect of inferential communication is that it presupposes  \emph{competence acquisition}. How do we internalise mental categories and build conceptual metaphors to process everyday events? We have introduced an epistemic difference between standard and non-standard events, but not determined how they relate, or how do we build abstract representations (like `furniture') from specific examples (like `this table', or `that chair').

How do we go from \emph{standard} events, i.e., from experiencing concrete events associated to one or several predicates ($\exists^{st} P(a)$), to a predicate \emph{in general} ($\exists P(a)$), which generalises across concrete and hypothetical events? We break down this learning process in three cognitive processes: %can express this idea as:
  
\begin{principle}
  We denote the process of \textbf{Idealisation (\emph{I})} as `$\overset{I}{\Longrightarrow}$', and define it as: 
\begin{equation}
 \Big[  \forall^{ st fin} y' \exists x  \ \ \Big/ \ \  \forall y \in A(y') \Big] \overset{I}{\Longrightarrow}  \Big[  \exists x \forall^{st} y \in A(y) \Big]
\end{equation}
\end{principle}

In the previous $ \forall^{ st fin}$ reads \emph{for all standard finite}. The intuition is that if a cognitive system has a property that applies to all standard finite examples to which it has been exposed, then it is possible to implement a process, which we call \textbf{Idealisation}, that makes this predicate apply  to \emph{any} standard examples (thus, the ones seen, but also possible examples not yet seen). 
Of course, this occurs spontaneously in humans and other animals. However, the implementation of such a generalisation mechanism in artificial systems is a central problem of machine learning: finding a function that fits best the training set but also generalises to stimuli beyond the training set is the main challenge when training artificial neural networks in classification problems.

In the previous expression $x$ stands for the context in which such an idealisation process occurs. The main idea is that this idealisation will only work in some contexts, not all of them.  
This highlights the importance of contextual factors when associating predicates to events. Indeed, contextual factors are of paramount importance in inferential communication. One of their main roles is to help determine subsets of the domain of a predicate.
 We can express this idea as follows:

\begin{principle}

 We denote the process of  \textbf{Selection  (\emph{S})} as `$\overset{S}{\Longrightarrow}$' and define it as:

\begin{equation}
\forall^{st} X \exists^{st} Y \forall^{st} z  \Big[ z \in Y \overset{S}{\Longrightarrow} z \in X \cap Y(z) \Big]
\end{equation}

\end{principle}

 The intuition behind the \textbf{Selection} process is that, given a predicate $Y$ associated to standard events, we can implement a cognitive process, which we call \textbf{Selection}  that, given  any predicate $X$, is able to determine which subset of events $z$ satisfy  predicate $Y$ and also $X$.
 
 If we think of properties $X$ and $Y$ as implemented in separate GCMs (which we can call $GCM_X$ and $GCM_Y$), this principle can be implemented at least in two ways. First, with the  output of $GCM_X$ feeding the inhibitory inputs $p$ of $GCM_Y$, and therefore affecting the domain on which $GCM_Y$ responds to stimuli consistent with predicate $Y$. Second, introducing a third module that processes the outputs of both $GCM_X$ and $GCM_Y$ and responds only to the combination of both.  However, implementing a separate module for \emph{any} combination of predicates that have some domain overlap can produce a combinatorial explosion. Therefore, to implement this principle it seems more efficient to opt for contextual influence of GCM modules that have domain overlap, despite the dynamics of contextual influence will be more challenging to understand.

The third cognitive principle  extends the domain of predicates from \emph{standard} events to events in general:
 
\begin{principle}

 We denote the process of \textbf{Transference (\emph{T}) } as `$\overset{T}{\Longrightarrow}$'  and define it as follows:
\begin{equation}
\Big[ \forall^{st} y A(y) \Big] \overset{T}{\Longrightarrow} \Big[ \forall y A(y) \Big]
\end{equation}

\end{principle}

The intuition behind the \textbf{Transference} principle is that if in a cognitive system there is a predicate that applies to all standard members, then we can implement a cognitive process, which we call \textbf{Transference}, by which we can make the predicate apply to all members. This mechanism allows to transfer the properties of standard members to all members. For example, to hypothetical sequences of events.

This undeniably happens in natural reasoning, where we learn to make inferences on hypothetical situations and events, and we even build conceptual metaphors to enrich the predicates associated with factual or hypothetical events. However, current artificial cognitive systems do not learn to generalise a predicate beyond the domain of standard examples on which it is trained. In artificial systems it is easy to simulate standard events using stored media to be presented as stimuli to a machine learning algorithm. 
However, it is difficult to see how we may implement in an artificial agent the notion of \emph{conceptual metaphor} to interpret the events involved in everyday communication. To do so, we would first need to clarify what are, in an artificial system, events that exist but do not have a standard existence. We can however assume such agents can be built, and explore their properties:

\begin{definition}
A \textbf{competence acquirer} is an  agent where  the 3 cognitive processes
\textbf{I} ($\overset{I}{\Longrightarrow}$), \textbf{S} ($\overset{S}{\Longrightarrow}$) and \textbf{T} ($\overset{T}{\Longrightarrow}$) are implemented. 
\end{definition}

\subsection{The consistency of inferential agents}
\label{maths}

If we implement artificial agents that have inferential skills and are also capable of  competence acquisition, will they develop consistent representations? We believe this is not the case in general, in the sense that \emph{any} inferential agent will develop consistent representations. However, here we show that a particular subset of inferential agents can develop consistent representations.

\begin{definition}
A binary agent is one where all the processes are binary rules. This applies to the transference functions in the GCM modules that form it and also, if implemented, to the three cognitive processes for competence acquisition previously defined.
\end{definition}

\begin{theorem}
There exists a binary inferential agent that is also a competence acquirer which is  a consistent embodiment of mathematical foundations.
\end{theorem}

More qualitatively, the way this theory is built also seems closer to an embodied account of mathematical foundations \cite{lakoff2000mc}:  numbers are defined as Frege numerals, which is not only the first introduced historically, it is also more intuitive: for example, the number 3 can be defined as a property shared by all sets which have 3 elements. This seems particularly appropriate here: if we are to build mathematical foundations consistently with how we build our internal representations, it seems reasonable that the definitions of notions for which we have an intuition should match such intuitions (intuitions which are, presumably, built from embodied representations). Similarly, in this set theory the universal set exists, the complement to a given set always exists, and it is possible to create any set simply through the union of two existing sets. In addition, with the 3 IST principles (see appendix) we can formalise calculus closer to how it was introduced historically, with infinitesimals, circumventing the need for definitions based on limits, and stating closer to the, arguably, intuitive notion of change at arbitrarily small increments. It can also be used to formalise probability theory in a much more simple way than the usual way, based on complicated results from measure theory \cite{nelson1987rep}.

\section{Discussion}
\label{discuss}

\subsection{Inferential communication}
The publication of  \emph{A mathematical theory of communication} \cite[]{shannon48} spread the view of communication as the transmission of a message encoded by an emitter and decoded by a receptor. 
Hardly ever a theory developed for engineering purposes has had such influence in fundamental research fields such as physics, linguistics, biology and cognitive sciences.  However, people studying natural communication started to raise some objections to such view of communication and they eventually proposed the inferential communication model \cite[]{sperber1986relevance}  to give some account of aspects of everyday conversations not described by classical information theory. The benefit of such a theory 
is that it allowed to explain why communicators convey much more information with their utterances than what is contained in their literal expressions. We have introduced a formal model that addresses this challenge and hints at some possible ways to engineer practical solutions to such interaction scenarios.

\subsection{Model summary}

In inferential communication an agent is an entity which processes information. This information is received as a stream of events. \emph{Events} are mainly actions of other communicative agents,  for which we assume such events are intentional actions, i.e., events generated by an agent to which other agents associate intentions.
However, part of the events can also be natural events that may be perceived as relevant for communication purposes. In both cases, when they receive information in the form of events, inferential agents associate  an interpretative meaning to these events.  This cognitive process, that we call \emph{interpretation}, consists in associating the events with communicative intentions, within a context that includes  a given situation, and a cultural background assumed as shared among the different agents. It can, sometimes, provoke agents to perform additional actions, either as an immediate reaction to them, or as a delayed answer.

We have argued this picture can be formalised in MaTIC: within the space of \emph{Language games}, communication is a collaborative game where each agent has a set of goals, some of which are shared. Grice's cooperative principle can be formalised as a goal shared among the participants in the game. The intentions of individual agents can be formalised as a set of goals. The events defining  the game are communicative actions, and are intended to be perceived, and associated with predicates. 

In MaTIC the idea of a standard event is associated with an epistemic definition of event, but also with a rigorous logical treatment. The particular predicates associated with an event are selected according to the general principles of relevance theory, with information theoretic measures. These associations are learnt  through a combination of adaptive GCM modules, together with connectivity changes guided by the I, S and T principles.  The existence of a property  is associated with the output of a cognitive module. Predicates are properties organised in  non-circular connectivity, something that allows building richer representations. 

In this picture, cognitive agents infer causal implicatures for the occurrence of an event $e$ by, first, finding the conditions in which $e$ \emph{can} occur and, second,  finding a motivation for $e$ to occur. In other terms: the context determines the \emph{possibility} of $e$. Correspondingly, the event or intention that determines the \emph{necessity} that $e$ occurs, constitutes the cause that we associate with it. The way a cognitive agent will do so will be based on heuristic reasoning, under general principles of uncertainty minimization.

The main difference between cognitive agents involved in inferential communication with the other existing systems introduced is that in those examples of GCM instances 
 the inhibitory input $p(t)$ was either null, or either it simply restricted the response of a system to a subset of possible outcomes. In addition, the inhibitory input function was determined \emph{beforehand}, i.e., before any excitatory input occurred. This does not seem appropriate to address the dynamic interplay that occurs in inferential communication.  In the conversation example previously introduced, the occurrence of an event such as `\emph{B says: There is a garage around the corner}' was, arguably, made possible by a larger context, such as B knowing this fact, as well as the previous events. In addition, assuming none of the participants had hidden intentions, an event such as \emph{A says: I am out of petrol} would be generally interpreted as the cause of B's statement.

More generally, as new exchanges occur in a conversation, the context of the conversation will evolve, and therefore the set of possible actions will too. \emph{Contexts in natural communication are dynamic}, therefore we need to consider GCM instances with response functions which combine the $p$ and $n$ inputs in a more intricate way. Overall, MaTIC does not define completely how causal implicatures occur, but it does define a set of \emph{requirements} that must be satisfied by the agents exchanging intentional actions. Under these requirements, and contrary to classical information theory, the probability of occurrence of an event changes across time.

\subsection{A neuroscience perspective}

Is MaTIC based on reasonable assumptions, from which we can build further details? Contemporary cognitive neuroscience suggest there may be empirical support for a computational theory of inferential communication. The mirror system is assumed to enable imitation, and empathy, and to be based on sensor-motor neural mechanisms   \cite{kohler2002hearing}. Such bodily disposition would explain the skill of learning other's intentions \cite{barron2020neuronal}. In fact, embodied and enactivist approaches to cognition are connecting with reasonable plausibility the spheres of bodies, societies, and cultural manifestations, with (inferential) communication as a multi-modal skill \cite{gallagher2018active}.

The modern study of intentionality 
encompasses both mental states towards things as well as the ways by which human agents understand others intentional dispositions \cite[]{brentano2014psychology}. This atonement process is located in our neural system, working from very basic sensor-motor level to symbolic processing \cite[]{rizzolatti2007mirror}, \cite[]{bretherton1991intentional}. For such reasons, there are mechanisms by which such intentional actions can be observed and therefore the related actions can be inferred to. In this sense, the study of intentionality allow us to predict and understand the next actions of human agents, allowing us to design a suitable answer. Indeed, in the model we propose agents will also, sometimes, perform additional actions as a reaction to the events and the predicates associated to them. According to this theory, this should be done in line with the first principle of relevance theory, i.e., the fact that \emph{human cognition tends to be geared towards the maximisation of the cognitive effect}.

There are other accounts on how abstract representations can be built through the combination of bottom up and top down processes. 
Notably, predictive coding and active inference  \cite[]{constant2020representation,friston2007free}, suggest information-theoretic measures are relevant to account for how the brain learns to interact with its environment. 
Despite MaTIC is compatible with predictive coding  and the idea that uncertainty minimisation can go quite far in determining cognitive processes, there  are two aspects that make MaTIC  strikingly different: first, in how it reconciles the idea of embodied reasoning with a formalism for mathematical foundations, and second in how it seems compatible with cognitive accounts of everyday exchanges, as conceptualised in cognitive pragmatics.

\subsection{Cognitive Pragmatics}
 
The process of \emph{interpretation} has also been interpreted in a Bayesian framework 
\cite{vallverdu2015bayesians}. 
 We believe the model introduced here may complement Bayesian models  of thinking processes by taking into account  the other embodied variables present into communicative processes. One of the key assumptions of Bayesian inference is that a belief and its opposite (in our notation, $A$ and $\neg A$) exclude each other: they form a complementary set. However, this is not a general assumption, for example, in deep learning inference systems, where Bayesian models are just a subset of the solutions that, in practice, work. 
In this context,  local restrictions may explain why the agents opt for selecting some informational sets and processing models instead than others: cultural and training/academic variables are fundamental thresholds of the  way of dealing with information 
(see the notion of blended cognition \cite{vallverdu2019blended}).
 Future work should analyse whether this model does different predictions than Bayesian modelling, and which gives better support to the analysis of the mental mechanisms that are used to process intentional communication.

In this context, we also point to Rational Speech Act Theory  \cite[]{yuan2018understanding}, which provides a fully formalised framework for pragmatic inference under a Bayesian framework. For example, \cite{goodman2013knowledge} shows that pragmatic agents do inferences that seem more adjusted to reality than their literal counterparts. Beyond its insistence on maximising utility, Rational Speech Act Theory seems  consistent with a Bayesian picture, and is reasonably aligned  with general principles related with reducing uncertainty in information exchanges, close to the idea of minimising surprisal, as discussed in the previous section. However, beyond the question of whether Bayesian modelling covers the whole picture, there are two aspects of MaTIC that do not seem addressed in the Rational Speech Act Theory: first, how such inferences can be supported by an embodied mechanism (for which we have proposed the GCM, and shown how biological and artificial systems can be seen as particular instances of such a modular embodiment) and, second, how the internal representations supporting pragmatic inference can be developed and maintained through time.

\subsection{Embodied reasoning and representation consistency}

A particularly original aspect of MaTIC is its link with formal set theory. In this context it is striking that actual infinity is not assumed. Instead, we have assumed the existence of external events, and predication, and argued we can build most of the usual construction for mathematical engineering (sets, numbers, vectors, functions, metrics, derivatives) from it.

If our model of competence acquisition is correct, the mechanisms that we posited for acquiring communicative competence should also apply to acquiring mathematical competence. However, this is at odds with how the foundations of mathematics are generally assumed by mathematicians. For example, in \cite{cohen1966set} we find an exposition of the default set theory, also called Zermelo-Frankel with Choice (ZFC). Such a theory assumes there only exist sets, and that there exist an actual infinite amount of them.
This is absolutely at odds with the notion of an agent developing internal representations through interactive sensor-motor loops in a finite world. In this context, it is particularly striking that our theory does not assume the actual existence of an infinite number of sets to build solid mathematical foundations. Rather, it is enough to exist the existence of internal and external events, and GCM modules processing them.

Future work should explore if MaTIC can also can be used to build agents that develop consistent internal representations from data, beyond the domain of mathematics. For this purpose, we believe again virtual reality provides an ideal experimental environment, one where artificial agents can easily build internal representations from their virtual environment. This would strongly simplify the perception challenges generally involved in robotics, and allow exploring directly whether agents in an environment can build internal representations of it, and whether this derives in exchanging information under pragmatic principles.

\subsection{A falsable theory}

 Any good scientific model must do accurate predictions that can be validated experimentally. The main implication of MaTIC might be the idea that by building artificial agents that satisfy the three cognitive principles introduced, then using data and machine learning these agents should be able to achieve inferential communication as competently as native speakers, and generate responses appropriate to their behaviour and also to their own communication intentions. We believe the best way to show this is through the iterative development of VR scenarios involving social interaction between autonomous agents and humans, based on these premises. 
 
 Another aspect of a scientific model is that it should be falsable. There are several ways to show this theory is wrong. For example, by:
 \begin{enumerate}
     \item showing that the three cognitive principles introduced are not relevant for the purpose at hand.
     \item showing that massive modularity does not occur, after all, in animals nor humans, or that it is not needed to build artificial systems that are capable of inferential communication. 
     \item showing that predication can be built with circular systems, that there is no difference between circular and non-circular systems
     \item showing that contextual processing does not occur through inhibitory inputs of GCM modules.
 \end{enumerate}

 Further computational systems could also be analysed within MaTIC. Here we are not short of options: Sperber has often stated that relevance theory is actually a general model of cognition. Since MaTIC is based on formalising some aspects of this theory, and on how to embody it, there is ground for challenging this theory with virtually any modelling approach used in systems neuroscience or in artificial intelligence focused on machine learning. 
 A separate stream of work analysing interpersonal interaction using methods derived from the modelling of dynamic physical systems that should be explored\cite[]{
 zhai2016design, alderisio2017interaction, lombardi2019deep}. 
 Future work should also focus on validating MaTIC in behavioural studies of people interacting with agents embodied in interactive virtual characters.
 Another way to explore whether MaTIC does valid predictions  is to to review  insights from behavioural economics and cognitive psychology on decision making. The discounting effect and similar paradoxes have been well characterised in behavioural economics, but are difficult to explain with classical probability theory. We may study whether an agent learning on these premisses will also adopt heuristic decisions that reflect these cognitive biases.  
Another direction to explore is to clarify whether MaTIC gives a better account for inferences and predicates as they occur \emph{in the world}, i.e., consistently with how cognitive psychology characterises natural categorisation and how cognitive pragmatics argues we reason and make decisions. From a developmental perspective: do the interpretation of the I, S and T principles introduced capture part or all of the picture in Vyggotsky's concept of \emph{Internalisation}?  These topics are being addressed by ongoing neuroscientific work \cite[]{teufel2020forms}, and it would be interesting to explore whether our theory is compatible or helps explain how such developmental processes work. Empirically, we could also look at the extent to which artificial agents developing their skills under such principles incur in phenomena such as  "hyper-regularisation", i.e., the trend kids to learn general rules for syntax, and then have to re-learn exceptions to the rule that they had initially learnt. 

\subsection{Ethics}

Last but not least, a different  question to consider is what ethical implications should be addressed if we were to implement such systems in practical, functional, consumer products. The arrival of immersive virtual and augmented reality to the consumer market, is likely to  have a considerable social impact particularly if combined with autonomous agents capable of everyday pragmatical  inferences as humans do everyday spontaneously and often unconsciously.  Depending on how it is implemented it can provide opportunities or challenges  to the realisation of human rights  in society. Let's introduce a simple example: the always winning Janken robot from Ishikawa Oku Lab (Japan), which has absolute winning rates playing at rock-paper-scissors against humans \cite[]{katsuki2015high}. The robot is able to capture and interpret bodily movements of the human player and then react more quickly always beating the human, which has not been able to understand the unfairness of the situation, because the robot moves itself following a similar temporal pattern than that of humans. This mechanism applied to other contexts would make possible to open a new way of cheating humans in benefit of machine owners. 
Is it reasonable to allow  these super-human affordances to be used without the users' knowledge? \cite[]{vallverdu2016emotional}. 
What considerations should be taken into account when our governments approach these topics?

\bibliographystyle{apacite}

\setlength{\bibleftmargin}{.125in}
\setlength{\bibindent}{-\bibleftmargin}

%\bibliography{CogSci_Template}

\bibliography{refdatabase120209}

\clearpage

\appendix
\section{Theorem proofs}

\subsection{Proof of theorem 1}

Theorem 1 states that \emph{The outputs of all non circular cognitive systems are Sets}. To prove this, we need three additional definitions:

 \begin{definition}
A statement is  \emph{stratified} if, and only if, when a variable appears in both sides of a membership statement ($\in$), then the right side of the expression is  inside one more level of brackets. For example,  `$ x \in [x,y ]$' is a stratified statement. Opposed to this, a  statements like $x \in x$ 
is \textbf{not} stratified. 
 \end{definition}

\begin{definition}
A cognitive system has a circular connection if, when following one of the outcomes  it is possible to reach an input.
\end{definition}

\begin{definition}
We consider the relation between the input and the output of a non-circular cognitive system to be stratified. To denote this fact, for a given input $a$, the output $A$ can also be expressed as $[a]$.
\end{definition}

We only consider stratification for non-circular cognitive systems because in  a circular connection, it will not be possible to define stratification in a non-ambiguous way. If we consider a trivial circular cognitive system, where the output is connected to the input, it is clear why an input $a$ and an outcome $[a]$ cannot be distinguished: an event $a$ cannot be consistently defined, independently from the properties $A$ associated with it. On the contrary, if the cognitive system does not have a circular connection between its inputs and its outputs, the relation between $a$ and $[a]$ is clearly defined: if the first corresponds to the input of the cognitive system, the second corresponds to its output. We therefore need to impose non-circularity to be able to guarantee stratification in a cognitive system.

Another important property of considering outputs of non-circular cognitive systems as sets is that inputs of non-circular cognitive systems do not only process outputs of other non-circular cognitive systems. They also process inputs of circular cognitive systems and inputs external to the entire agent of which they form part. Therefore, if we consider the outputs of  non-circular cognitive systems as sets, we must also consider the existence of non-sets, also known as urelements.
It is a remarkable result of alternative set theory that if we only consider stratified statements,  and the existence of urelements, then we have a set theory that is consistent.
This result was introduced in 
\cite{jensen1968csm}, and is explained more pedagogically in \cite{holmes1998elementary}. A pedagogical introduction explaining the role of urelements and of weak extensionality can be found in \cite{holmesStanford}. For our case of interest it implies that, in non-circular cognitive systems, \emph{all properties are sets}. Of course, the previous is assuming the fact that the outcome of a cognitive system may not be binary. However, the generalisation from binary systems is straightforward, we only need to generalise the definition of a set to include \emph{fuzzy} sets.

\subsection{Proof of theorem 2}

How does a cognitive system determine an implied cause? 

%For a given event $e$, there is always a possible  cognitive system that can determine the most likely implied cause using an information-based metric such as surprise minimization.

Lets consider a cognitive system in a space of events. For a given event $y$ there will necessarily be some events that have happened before it. Lets consider an arbitrary subset of the preceding events $y^*$, the context of $y$. Lets also consider an arbitrary element preceding $y$, and distinct from the context $y^*$, as the cause $y^{**}$. If we build a GCM where the occurrence of the context $y^*$ determines $p(t)$, and $n(t)$ is determined by the occurrence of $y^{**}$, then the output $o(t)$ may be the estimated likeliness that $y{**} \Rightarrow y$. 
We can build a cognitive system made of different GCM modules, each considering a different context and a different cause (always, among preceding events). With different occurrences of the event $y$, the GCM module that most often predicts the occurrence of $y$ from preceding events will be the one that will minimize surprisal.

%For  a  given  event  e,  a  cognitive  system  canalways  establish  the  most  likely  implied  cause  using  aninformation-based metric such as surprise minimizatio

More technically, we are using  the notion of lattice \emph{pseudo-complement}, as defined in Heyting algebras \cite[]{rutherford1965ilt}. 
%Going into further detail, the first step for such a GCM is to determine the possibility of an event, i.e. whether the context makes this event possible. %JOAN add REFS PSYCHOLOGY.
If, among all the events preceding $y$  we consider only  its cause $y^*$ and the events forming its context $y^{**}$, we notice that we can associate a semi-lattice to any $y$. In this construct, the implication operator ‘$\Rightarrow$’ corresponds to the definition of implication in a Heyting algebra \cite[]{rutherford1965ilt}.  We can always build such a construct: by definition a cause and a context always precede the event, and by definition a pseudo-complement is unique.  In this formalism implying a cause for an event is equivalent to build such construct, and therefore it will always be possible to build a cognitive system to infer  a causal implicature: $(e^{**} \Rightarrow e)$.

\subsection{Proof of theorem 3}

In classical information theory communication is a \emph{stationary stochastic} process. A stochastic process is, by definition, a sequence of events whose occurrence is determined by probabilities. \emph{Stochastic} means that each communication event $x$ has a probability of occurrence $p(x)$  associated. \emph{Stationary} means and that this probability does not change through time ($p(x,t) \equiv p(x)$). For example, the probability of an event such as \emph{receiving a letter `g'} in a telegraph is constant through time. Opposed to this, in inferential events occur in a communication scenario where the basic building blocks are intentional actions, and in this context it is not possible to consider communication as \emph{Stationary}.   
To show in this scenario communication is not stationary, the easiest is to use the classical definition of information:

\begin{equation}
 H(x) = \sum_{i=1}^N p(x_i) log p(x_i)
\end{equation}

In classical information theory the set of possible symbols $1..N$ does not change through time. However, in inferential communication  the set of possible symbols is determined by the context. For example, in a given context it is possible that a person says a sentence like \emph{do you want to marry me?}, and we can associate a given probability to this event. However, in other contexts this is not a possible sentence, and therefore its probability is 0 (since by construction  $p(x) \leq pos(x) $). In this technical sense this theory seems  more general than classical information theory: we do not assume communication processes to be \emph{Stationary}. This means that the probability associated to an event may change through time, as well as the information associated with its occurrence. 

Closer to a communication scenario: an inferential agent will perform an intentional action depending on his own goals. Which  action most contributes to satisfy these goals will be strongly dependent on the context in which it is performed, and the context will change constantly as the different agents perform different intentional actions.

\subsection{Proof of theorem 4}
The proof of consistency is a constructive one, and one that draws extensively from existing results.  
%In theorem 1 we showed how non-circular cognitive systems could be seen as an embodiment of set formation, when we define the idea of a Set from the notion of Stratification.
Here we only consider binary outputs, which means the sets will always be crisp sets (i.e., not fuzzy). The set theory based on stratification with the existence of urelements is called NFU.  Holmes \cite[chapters1 to 11]{holmes1998est} shows how to define relations, and functions. 
Even more, in chapter 12 it is shown how Frege numerals, which were the initial definition of numbers in set theoretic terms, work in this set theory, and fail in the much more often adopted ZFC.% (as opposed to Von Neuman numerals). 
However, NFU cannot prove the existence of infinity. In this sense, NFU is weaker than Peano arithmetic. %As a consequence,  it is impossible to prove the existence of real numbers,the continuum, etc. 
And introducing the existence of infinity as an axiom does not fit with the idea of developing mathematical foundations based on learning embodied representations.
To clarify how this occurs in our theory, we turn towards the binary versions of the principles previously introduced:

\begin{enumerate}
    \item \textbf{Idealisation}
 \begin{equation*}
 \Big[  \forall^{ st fin} x' \exists y  \ \ \Big/ \ \  \forall x \in A(x') \Big] \Longleftrightarrow  \Big[  \exists y \forall^{st} x \in A(x) \Big]
\end{equation*}

    \item \textbf{Selection}
\begin{equation*}
\forall^{st} X \exists^{st} Y \forall^{st} z  \Big[ z \in Y \Longleftrightarrow z \in X \cap Y(z) \Big]
\end{equation*}
 
 \item \textbf{Transference}
\begin{equation*}
\Big[ \forall^{st} y A(y) \Big] 
\Longleftrightarrow \Big[ \forall y A(y) \Big]
\end{equation*}

\end{enumerate}

\cite{nelson1977ist} showed that the previous three principles, when applied over ZFC set theory (the most-studied, assumed by default, set theory), constitute a conservative extension of such a theory, which we call IST. In the context of set theory, the particularity of IST is that it introduces `\emph{standard}', a predicate that applies to formulas. It allows statements like \emph{ $x$ is \emph{infinitesimal} iff for all standard $\epsilon > 0$, we have $|x| \leq \epsilon$.}, or \emph{ x is limited in case for some standard $r$ we have $|x| \leq r$
}. 
It is important to highlight the predicate \emph{standard} is not based on set membership. Therefore, one cannot form a subset with an \emph{external} formula, it is \emph{illegal set formation}.  However, this does not imply that a formula with a \emph{standard} predicate cannot be associated with a set. It merely means that one must remove the predicate `\emph{standard}' from the formula in order to find what set it corresponds to.  The rules to do so are the 3 axioms of IST.
%The intuition behind the \textbf{T} principle is that if a particular formula applies to all standard cases, it applies to all cases. By successive applications of \textbf{(T)} one can show that $\forall^{st} A \Leftrightarrow \forall A $. The converse is also true: to prove an internal theorem it is enough to show it works for all standard cases. However, when applying the \textbf{T} principle one has to be careful with illegal transfer.  
An example by Nelson explaining the \textbf{Transference} principle:
\begin{quote}
As an example of transfer, we know that for all real $x > 0$  there is a natural number n such that $n*x \geq 1$; therefore, for all standard $x > 0$ there is a standard $n$ such that $n*x \geq 1$. But suppose that $x > 0$ is infinitesimal. Do we know that there is a natural number n such that $n*x \geq 1$?  Of course; we already know this for all $x>0$. But if we try to argue as follows --``there is an n such that $n*x \geq 1$; therefore, by the dual form of transfer, there is a standard n such that $n*x \geq 1$''-- then we have made an error: transfer is only valid for the standard values of the parameters (in this case x) in the formula. This is an example of illegal transfer.
(...)
Another form of illegal transfer is the attempt to apply it to an external formula. For example, consider ``for all standard natural numbers n, the number n is limited; by transfer, all natural numbers are limited''. This is incorrect. Before applying transfer, one must check two things: that the formula is internal and that all parameters in it have standard values. 
\end{quote}

However, Nelson applied these theorems over ZFC, and here we are considering these over 3 principles in the context of NFU.

\begin{lemma} \textbf{There exists a non-standard object }.
\end{lemma}

The proof is introduced as an example in \cite{nelson1977ist}:
\emph{
Let $\phi$ be the formula $y  \not\equiv x$. Then for every finite set $X$, and so in particular for every standard finite set $X$, there is a $y$ such that for all $x$ in $X$ we have $y  \not\equiv x$. Therefore, there exists a nonstandard $y$. The same argument works when x and y are not restricted to range over a finite set. 
}

The previous lemma shows that given any finite set A there exists a non-standard object which does not belong to A. We now define for any set $A$ the incremental set $A + 1$ and the collection of inductive sets.
\begin{definition} The \emph{ incremental set} is the union of the members a of A and singletons of elements not in a: $\{a \cup \{ x \} |  a \in A  \wedge x \notin a \}$
\end{definition}

\begin{definition} An \emph{inductive set} is one which contains 0 and contains the incremental set of each of its elements. The collection of inductive sets $Ind$ is: $\{A | 0 \in A \wedge A+1 \subseteq  A \}$
\end{definition}

with it is beyond any finite number. 

\begin{lemma} \textbf{There exist an infinite amount of natural numbers} From the definition of incremental set, it can be proven that there exists a set which has an \textbf{infinite} amount of members, in the sense that the number associated 
\end{lemma}

\begin{lemma} \textbf{The universe V is a standard predicate:  $\exists V  \Leftrightarrow \exists^{st} V  $}. The proof is done by stratified comprehension, $\exists V  \equiv \{x | x\equiv x \} $. 
\end{lemma}

\end{document}